\documentclass[%
reprint,
amsmath,amssymb,
aps,
prb,
]{revtex4-1}
\usepackage{graphicx}
\usepackage{dcolumn}
\usepackage{bm}
\usepackage{graphicx}
\usepackage{subcaption}
\usepackage{tabularx}
\usepackage{amsmath}
\usepackage{mathtools}
\usepackage{hyperref}
\begin{document}

\title{
How to accurately determine a saturation magnetization of the sample\\ in a ferromagnetic resonance experiment?
}

\author{P.~Tomczak} 
\email{e-mail: ptomczak@amu.edu.pl}
	\affiliation{Quantum Physics Division\\ Faculty of Physics, Adam Mickiewicz University 
ul. Umultowska 85, 61-614 Pozna\'n, Poland}
\author{H.~Puszkarski}
\affiliation{Surface Physics Division \\ Faculty of Physics, Adam Mickiewicz University 
ul. Umultowska 85, 61-614 Pozna\'n, Poland} 

\date{\today}

\begin{abstract}

The phenomenon of ferromagnetic resonance (FMR) is still being exploited for
determining the magnetocrystalline anisotropy constants
of magnetic materials. We show 
that 
one can also determine accurately the saturation magnetization 
of the sample using results of FMR experiments
after
taking into account the relationship between resonance frequency and curvature of 
the spatial distribution of free energy
at resonance.
Specifically, three examples are given
of calculating 
saturation magnetization from FMR data:
we use historical Bickford's measurements from
1950 for bulk magnetite, 
Liu's measurements from 2007
for a 500 mn thin film of a weak ferromagnet (Ga,Mn)As,
and 
Wang's measurements from 2014 for an ultrathin film of YIG.
In all three cases, the magnetization values we have determined 
are consistent with the results of measurements.
\end{abstract}

\maketitle
{\em \bf Introduction} ---
For a very long time, the ferromagnetic resonance has been 
used successfully to find magnetocrystalline anisotropy constants 
and spectroscopic splitting factors of ferromagnets, see e.g., Ref. [\onlinecite{Farle1998}].
Recently, we have shown [\onlinecite{Tom2018}] that using this classic experimental technique
completed with a cross-validation of the numerical solutions 
of Smit-Beljers equation [\onlinecite{Smit55, Artman1957, Wigen88, Morrish}], not only makes it possible 
to determine very precisely the spectroscopic splitting 
factor $g$ and magnetocrystalline anisotropy constants,
e.g., up to fourth order for the diluted magnetic semiconductor 
(Ga,Mn)As, but also to state which
anisotropy constants are necessary to properly describe 
the spatial distribution of the energy which is related to magnetocrystalline 
anisotropy. In what follows we present that 
one can interpret the results of FMR experiment in such a way
that it is possible
to determine the saturation 
magnetization of the sample under investigation, i.e, that it
is possible to find the spatial dependence 
of magnetocrystalline energy stored in a ferromagnet
using data collected in a single FMR experiment.

We will show how to determine saturation magnetization 
for three cases: for bulk magnetite, $\mbox{Fe}_3\mbox{O}_4$, we will use the historical Bickford's 
measurements [\onlinecite{Bickford1950}] from 1950, for a 500 mn film of a weak diluted ferromagnet, (Ga,Mn)As,
we will use Liu's measurements [\onlinecite{Liu2007}] from 2007 and for an ultrathin film of YIG 
we will use Wang's measurements [\onlinecite{Wang2014}] from 2014.
 
\begin{figure}[ht]
  \includegraphics[width=0.33\linewidth]{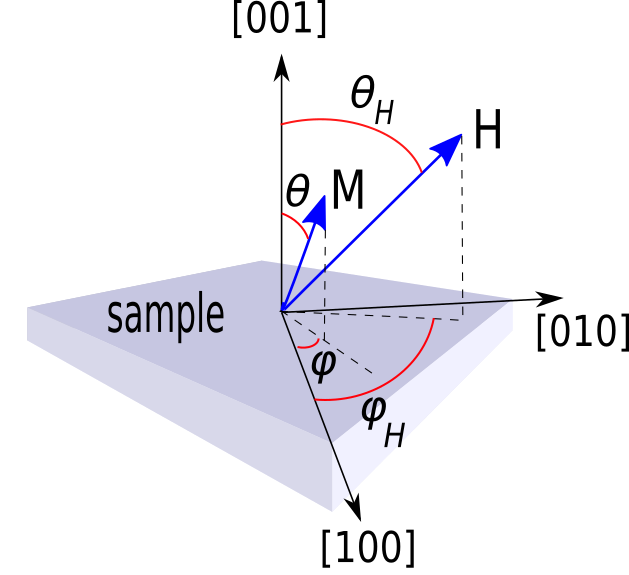}
  \caption{
  The geometry of the FMR experiment.
  The orientation of the applied
   magnetic field~${\mathbf H}$ is usually described in the
   coordinate system attached to the crystallographical
   axes of the sample. The field direction is represented by angles
    $\vartheta_H$ and $\varphi_H$ and 
    equilibrium direction of the sample     
magnetization~${\mathbf M}$ is represented by angles $\vartheta$ and $\varphi$.
}
  \label{sample_config}
\end{figure}
\vspace{0.5cm}
\noindent
{\em \bf What does one measure while performing FMR experiment?} ---
The most simple answer is that the spatial distribution of resonance field $\mathbf{H}_r$ is measured, i.e., 
its dependence on angles $\vartheta_H$ and $\varphi_H$, see Fig.~\ref{sample_config}.
However, let us consider two possible ways of performing the FMR experiment.
First, while fixing a frequency $\omega$  of an alternating microwave field 
one measures the dependence of the static magnetic field $\mathbf{H}_r$ on angles 
$\vartheta_H$ and $\varphi_H$ at resonance, 
and the second --- a static magnetic field is fixed and one changes the frequency of microwave field 
to find a frequency at which the precession of the magnetic moment of the sample occurs.
The resonance condition, for both measurement methods,  is given by the following Smit-Beljers equation
[\onlinecite{Smit55, Artman1957, Wigen88, Morrish}]
 \begin{equation}
\bigg (\frac{\hbar\omega}{g \mu_B} \bigg)^2 = \frac{1}{\mbox{sin}^2\vartheta}
(f_{\vartheta \vartheta} f_{\phi \phi} - f^2_{\vartheta \phi}),
\label{SB_eq}
\end{equation}
where $f$ is free energy of the sample divided by its saturation magnetization $M$,
i.e., free energy is expressed in terms
of real (Zeeman term) and fictitious (demagnetizing and anisotropy terms) fields. $f_{\vartheta \phi}=\frac{\partial f}{\partial \vartheta}\frac{\partial f}{\partial \varphi}$,
$g$ is the spectroscopic splitting factor,
$\mu_B$ -- the Bohr magneton and $\hbar$ -- the Planck constant.

Let us recall, 
that the Gaussian curvature of the surface $\mathcal{S}(\vartheta, \varphi)$ at one of its points,
being the product 
of two principal curvatures $\kappa_1$ and $\kappa_2$, is the determinant of the Hessian matrix
calculated at this point. One obtains in spherical coordinates 
 \begin{equation}
 \kappa_1  \kappa_2 = \mbox{det}
    \begin{bmatrix}
    \frac{\partial^{2}\mathcal{S}}{\partial \vartheta^2 } & \frac{1}{\sin \vartheta} 
    \frac{\partial^{2}\mathcal{S}}{\partial \mathcal{S} \vartheta  \partial  \varphi} 
    - \frac{\cot \vartheta}{\sin \vartheta}\frac{\partial \mathcal{S}}{\partial \varphi}\\
    \frac{1}{\sin \vartheta} \frac{\partial^{2}\mathcal{S}}{\partial \varphi\partial \vartheta } 
    - \frac{\cot \vartheta}{\sin \vartheta}\frac{\partial \mathcal{S}}{\partial \varphi}& 
    \frac{1}{\sin^2\vartheta}\frac{\partial^{2} \mathcal{S}}{\partial \varphi^2} +
    \cot \vartheta \frac{\partial \mathcal{S}}{\partial \vartheta}
  \end{bmatrix}.
  \label{curvature_eq}
\end{equation}
Let us now assume that we are interested in the curvature of the  
the free energy landscape of the sample at resonance and put $\mathcal{S}=f$
in Eq.~(\ref{curvature_eq}). 
Remembering that $\bf M$ precesses around its eqilibrium position, 
that is, the equations
\begin{equation}
\frac{\partial f}{\partial \vartheta}=0, \,\,\, \frac{\partial f}{\partial \varphi}=0
\label{equilibrium_eq}
\end{equation} 
are satisfied,
we arrive at Eq.~(\ref{SB_eq}) if we interpret $\big (\frac{\hbar\omega}{g \mu_B} \big)^2$
as a Gaussian curvature of the spatial distribution of free energy. 

Returning to the two ways of measuring the resonance field we see,
that by fixing a static field and subsequent measuring the frequency
$\omega$ of a microwave 
field for different angles at resonance, we see directly a spatial distribution 
of the curvature of the free energy landscape. 
If, however, we set the microwave field, 
and then by changing the static field
we hit the resonance ($\mathbf H_r$) for different angles $\vartheta_H$ and $\varphi_H$
then we will maintain a constant Gaussian curvature of the energy landscape during the measurement.

The answer to the question posed at the beginning of this section 
is, that while performing a typical FMR experiment by fixed $\omega$, 
we measure such a field $\mathbf H_r$ at which 
the Gaussian curvature $ \kappa_1 \kappa_2$ of the free energy landscape is constant, i.e., does not depend
on $\vartheta_H$ and $\varphi_H$. 
As we shall see,
this observation will not only allow finding contributions 
from various types of anisotropy to the free energy but also
will enable the determining of saturation magnetization of the sample.

\vspace{0.5cm}
\noindent
{\em \bf What determines the spatial distribution of the Gaussian curvature of the free energy at resonance?} ---
Assume that the free energy density $F(\vartheta_H, \varphi_H)$  of a ferromagnet placed in a magnetic field, 
expressed in terms of {\em magnetic} fields,
is given by
 \begin{equation}
 \begin{split}
   \frac{F(\vartheta_H, \varphi_H)}{M} =
   f(\vartheta_H, \varphi_H)  = 
  -H\sum_{\alpha} n_{\alpha}n^H_{\alpha} \\
  +M\sum_{\alpha, \beta}  \mathcal{N}_{\alpha \beta} n_{\alpha} n_{\beta}
  +\frac{1}{2} H_{c}\sum_{\alpha \ne \beta} n^2_{\alpha} n^2_{\beta}.
\label{Fict_Bickford}
\end{split}
\end{equation}
${\alpha, \beta}=x, y, z$.
The first term on the right-hand side of Eq.~(\ref{Fict_Bickford}) is the Zeeman energy, the second term 
is the demagnetizing energy, with $\mathcal{N}_{\alpha \beta}$
being the demagnetization tensor.
The 
last term describes the energy related to cubic magnetocrystalline anisotropy.
The components of vectors $n_{\alpha}^H=\mathbf H /H$ and $n_{\alpha}=\mathbf M/M$
are defined as usuall, see e.g., Eqs. (3a) and (3b) in Ref.~[\onlinecite{Tom2018}] and Fig.~\ref{sample_config}.
Note that with fulfilled relations (\ref{equilibrium_eq}), the free energy of a sample with a specific geometry and 
saturation magnetization in a magnetic field does not depend on angles $\vartheta, \varphi$ explicitly.

\begin{figure}[ht]
\begin{subfigure}{.11\textwidth}
   \includegraphics[width=0.9\linewidth]{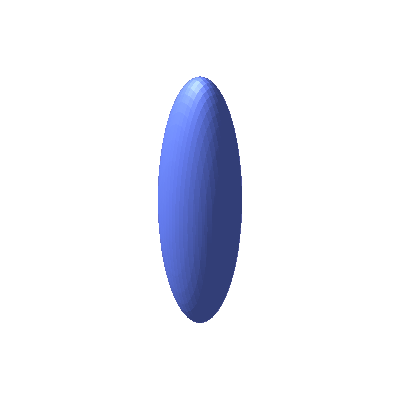}
  \caption{}
\end{subfigure}
\begin{subfigure}{.11\textwidth}
  \includegraphics[width=.9\linewidth]{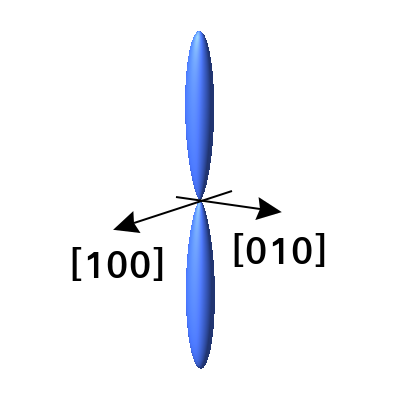} 
  \caption{}
\end{subfigure}
\begin{subfigure}{.11\textwidth}
  \includegraphics[width=0.9\linewidth]{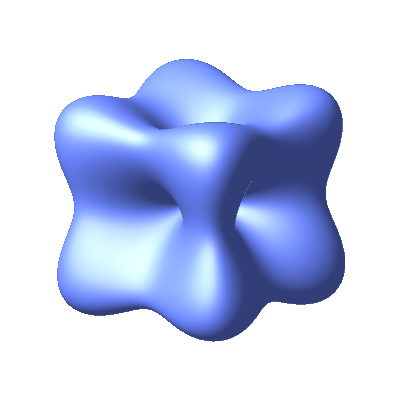}  
  \caption{}
\end{subfigure}
\begin{subfigure}{.11\textwidth}
  \includegraphics[width=0.9\linewidth]{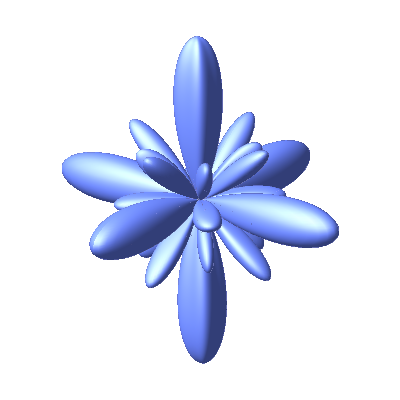} 
  \caption{}
\end{subfigure}
\caption{ 
  Terms in Eq.~(\ref{Fict_Bickford}) representing fictitious magnetic fields and their Gaussian curvatures
  (a)~$M\sum_{\alpha, \beta}  \mathcal{N}_{\alpha \beta} n_{\alpha} n_{\beta}$ (ellipsoid)
  (b)~its curvature
  (c)~$\frac{1}{2} H_{c}\sum_{\alpha \ne \beta} n^2_{\alpha} n^2_{\beta}$
  (d)~its curvature.
  }
\label{Bickford_basis}
\end{figure}

Each component of Eq.~(\ref{Fict_Bickford}) describes 
a spatially varying magnetic field with a certain curvature.
For example the spatial dependencies of the first two terms and their 
curvatures are shown in Fig.~\ref{Bickford_basis}.
Surface (a) - ellipsoid - represents the fictitious demagnetizing field 
for the sample oriented along axes [100], [010] and [001].
Its curvature is shown in Fig.~\ref{Bickford_basis}b.
Note that it is non-zero in the (001) plane, although this is not visible due to the used scale. 
The surface (c) represents the first order cubic magnetocrystalline fictitious field
and in Fig.~\ref{Bickford_basis}d is shown its curvature. 
The curvature of the Zeeman term in Eq.~(\ref{Fict_Bickford}) depends on 
the static magnetic field and can be changed during the measurement
while the curvatures associated with the fictitious demagnetizing and cubic anisotropy fields remain
constant during the measurement.

\vspace{0.5cm}
\noindent
{\bf How to find saturation magnetization and anisotropy fields numerically?} ---
During the FMR experiment the total Gaussian curvature of the free energy
remains constant: contributions of curvatures from various fictitious
magnetic fields (anisotropy, demagnetizing), related to the sample itself,
are compensated by the curvature of the Zeeman term.
To find numerically values of those fields, 
$g$-factor and saturation magnetization
we apply a procedure to some extent reverse to the measurement: 
for a given set of measured $H_r(\vartheta_H,\varphi_H)$ data, 
collected in a single experiment,
the constant $g$, $M$ and anisotropy and demagnetizing fields should be chosen so that
Eq.~(\ref{SB_eq}) should be met
for all measured static resonance fields $H_r(\vartheta_H, \varphi_H)$.
It means, that minimizing an appropriate objective function, which measures the deviation
from the constant curvature with respect to unknown anisotropy fields, $g$ and 
$M$ lets us find them.
This procedure is described in detail in Ref.~[\onlinecite{Tom2018}].
The key point for finding saturation magnetization is 
that its change leads to the change of the curvature of the demagnetizing field.
But to ensure that the determined magnetization value is unambiguous, the constraint
$\mathcal{N}_x+\mathcal{N}_y+\mathcal{N}_z=1$
should be met during the minimization procedure.

To carry out the minimization procedure, it is necessary to know the functional form of free energy.
Usually one assumes its specific form taking into account the symmetry of the system ---
free energy should be invariant 
under relevant symmetry
transformations.
Then it is possible
to expand it into basis functions (orthogonal or non-orthogonal)
with the same symmetry.
Typically this expansion is limited to some low-order terms 
of systematically decreasing basis functions.
It should be remembered, however, 
that in the case of examination of curvature, 
the higher-order basis functions
may give greater contributions to total curvature of free energy
than the lower-order ones.

\begin{figure}[!ht]
  \includegraphics[width=.8\linewidth]{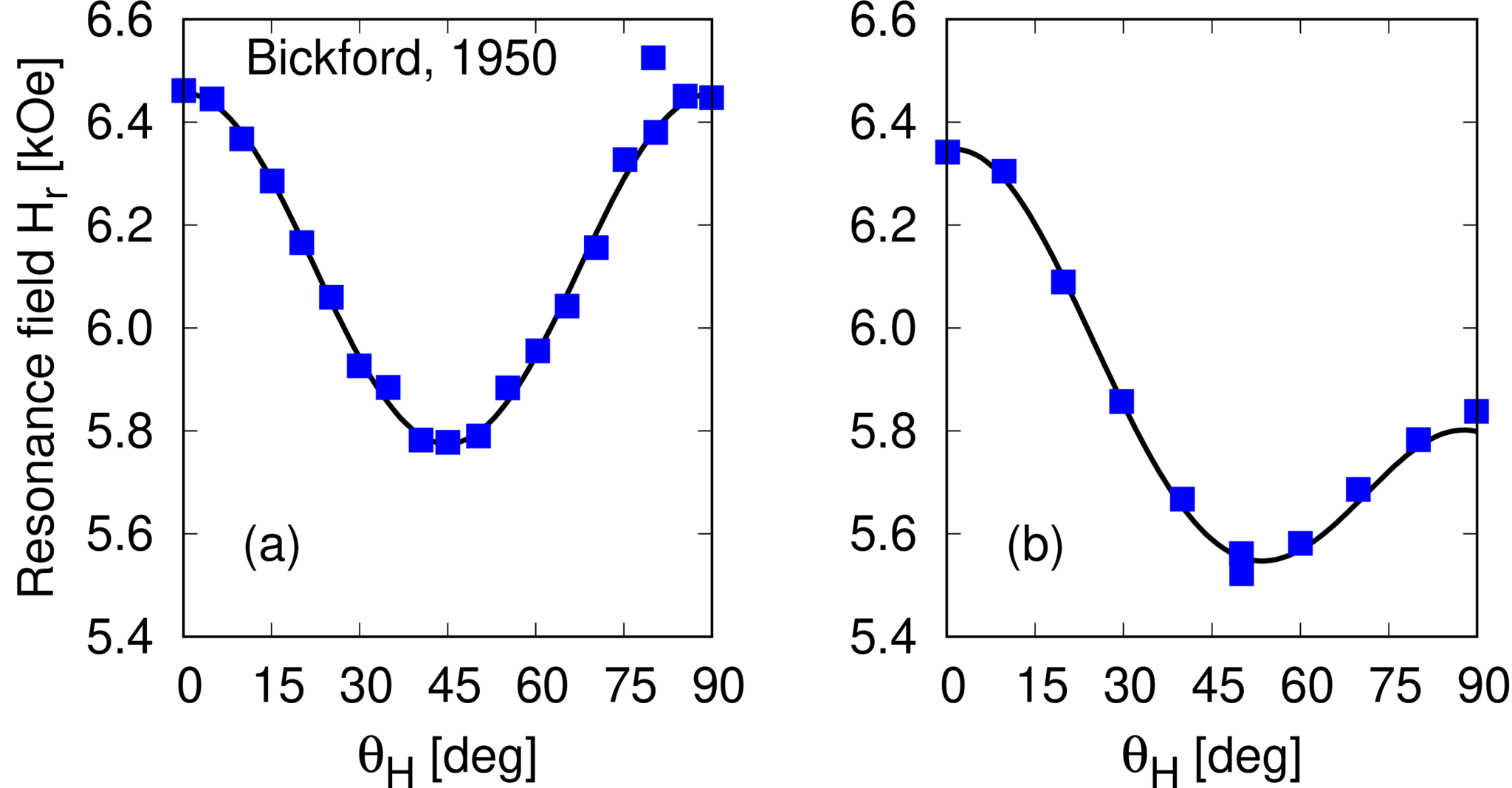}
   \caption{
  Resonance field versus out-of-plane angle $\vartheta_H$ for bulk magnetite for
  samples cut in (100) -- (a) and (110) planes -- (b).
  Squares - Bickford's measurements [\onlinecite{Bickford1950}], 
  black line - solution of Smit-Beljers equation for the vector ${\mathbf h_B}$
  given by Eq.~(\ref{nv_fields}).
  Figs (a) and (b) correspond to Figs~3 and 4 in Ref.~[\onlinecite{Bickford1950}], respectively.
  }
  \label{Bickford_resonance}
\end{figure}

\vspace{0.5cm}
\noindent
{\em \bf Numerical example I: Saturation magnetization of bulk magnetite} ---
Bickford [\onlinecite{Bickford1950}] measured the resonance fields for disk-shaped samples of magnetite 
cut in (100) and (110) planes.
The results of his measurements are shown as blue squares in Figs~\ref{Bickford_resonance}a and~\ref{Bickford_resonance}b,
respectively.
Each sample was oriented differently with respect to the crystallographic planes 
and thus it has different demagnetization tensor. We will assume, however, that both types 
of samples have the same $g$-factor and the same saturation magnetization. 
The demagnetizing
tensor is assumed to be diagonal for samples cut in (100) plane
\begin{equation}
\mathcal{N}_{(100)} = 
\begin{bmatrix} 
\mathcal{N}_x & 0 & 0\\
0 & \mathcal{N}_y & 0\\
0 & 0 & \mathcal {N}_z
\end{bmatrix}.
\label{N100}
\end{equation}

The curvature of free energy distribution, given by Eq.~(\ref{SB_eq}), depends now on six parameters 
which we denote collectively by the vector 
$
{\mathbf h_B} = (g, M, \mathcal{N}_x, \mathcal{N}_y, \mathcal{N}_z, H_{c}).
\label{v_fields}
$
Using the procedure described in Ref. [\onlinecite{Tom2018}], we get the following coordinates
of the vector ${\mathbf h_B}$
\begin{equation}
 \begin{split}
{\mathbf h_B} =  (
2.220(88),
5.906(150),
0.2216(267), 
0.2212(268),\\
0.5572(535),
-0.2392(95)).
\label{nv_fields}
 \end{split}
\end{equation}
Both, $M$ and $H_c$ are given in kOe. The errors (in brackets)
refer to the last significant digits 
and were calculated assuming that the 
measurement errors are subject to normal distribution.
Having determined the vector ${\mathbf h_B}$, 
we solve Eq.~(\ref{Fict_Bickford}) numerically 
with respect to ${\mathbf H}_{\mathrm{r}}$
and get the resonance field dependence on the $\vartheta_H$ 
angle (black line in Fig.~\ref{Bickford_resonance}a).
To consider measurements of samples cut in the plane (110)
let us rotate the coordinate system in such a way,
that the $z$ axis is perpendicular to the (110) plane.
The demagnetizing
tensor is given by
\begin{equation}
\mathcal{N}_{(110)}\!=\!
\begin{bmatrix} 
\frac{1}{4}(\mathcal{N}_{xy}+2\mathcal{N}_z)& \frac{1}{4}(\mathcal{N}_{xy}-2\mathcal{N}_z)&\frac{\sqrt 2}{4}(\mathcal{N}_y-\mathcal{N}_x)\\
\frac{1}{4}(\mathcal{N}_{xy}-2\mathcal{N}_z) &\frac{1}{4}(\mathcal{N}_{xy}+2\mathcal{N}_z)&\frac{\sqrt 2}{4}(\mathcal{N}_y-\mathcal{N}_x) \\
\frac{\sqrt 2}{4}(\mathcal{N}_y-\mathcal{N}_x)&\frac{\sqrt 2}{4}(\mathcal{N}_y-\mathcal{N}_x)& \frac{1}{2}\mathcal{N}_{xy}
\end{bmatrix}\!\!,
\label{N110}
\end{equation}
$\mathcal{N}_{xy}$ stands for $\mathcal{N}_{x}+\mathcal{N}_{y}$.
Solving again Eq.~(\ref{Fict_Bickford}) for such a tensor $\mathcal{N}_{(110)}$,
we reach the dependence of ${\mathbf H}_{r}(\vartheta_H)$ shown in 
Fig.~\ref{Bickford_resonance}b.
In Table~\ref{Bickford_Fields} the relevant anisotropy constants found by Bickford are presented
compared with those 
obtained in the way as described in the text.

To summarize this example, let us stress two things.
First, 
we observe a fairly large statistical errors
that are the result of the scattering of original measurements reported in Ref.~[\onlinecite{Bickford1950}].
Second, in fact, the Bickford's samples had the shapes of a circular flattened disks,
since in (100) geometry $\mathcal{N}{x}\approx\mathcal{N}{y}\ne\mathcal{N}{z}$.
\begin{table}
\caption{
Comparison of $g$-factor, saturation magnetization $M$ and cubic anisotropy
constant $K_{c}$ values
found by Bickford [\onlinecite{Bickford1950}]with those
obtained by the method described
in the text.
The errors (in brackets) refer to the last significant digits.
}
\begin{tabular}{cccc}
\hline 
                                  & $g$   &$M$ [kA/m] & $K_{c}$ [kJ/m$^3$]\\
\hline
                                  & 2.220(88) & 470(12)  & -14.12(92)\\
Ref. [\onlinecite{Bickford1950}]  & 2.07      & 472\footnote{ Measurement result, [\onlinecite{Bickford1950}].}      & -11       \\
\hline
\end{tabular}
\label{Bickford_Fields}
\end{table}

\begin{figure}[!ht]
 \includegraphics[width=.8\linewidth]{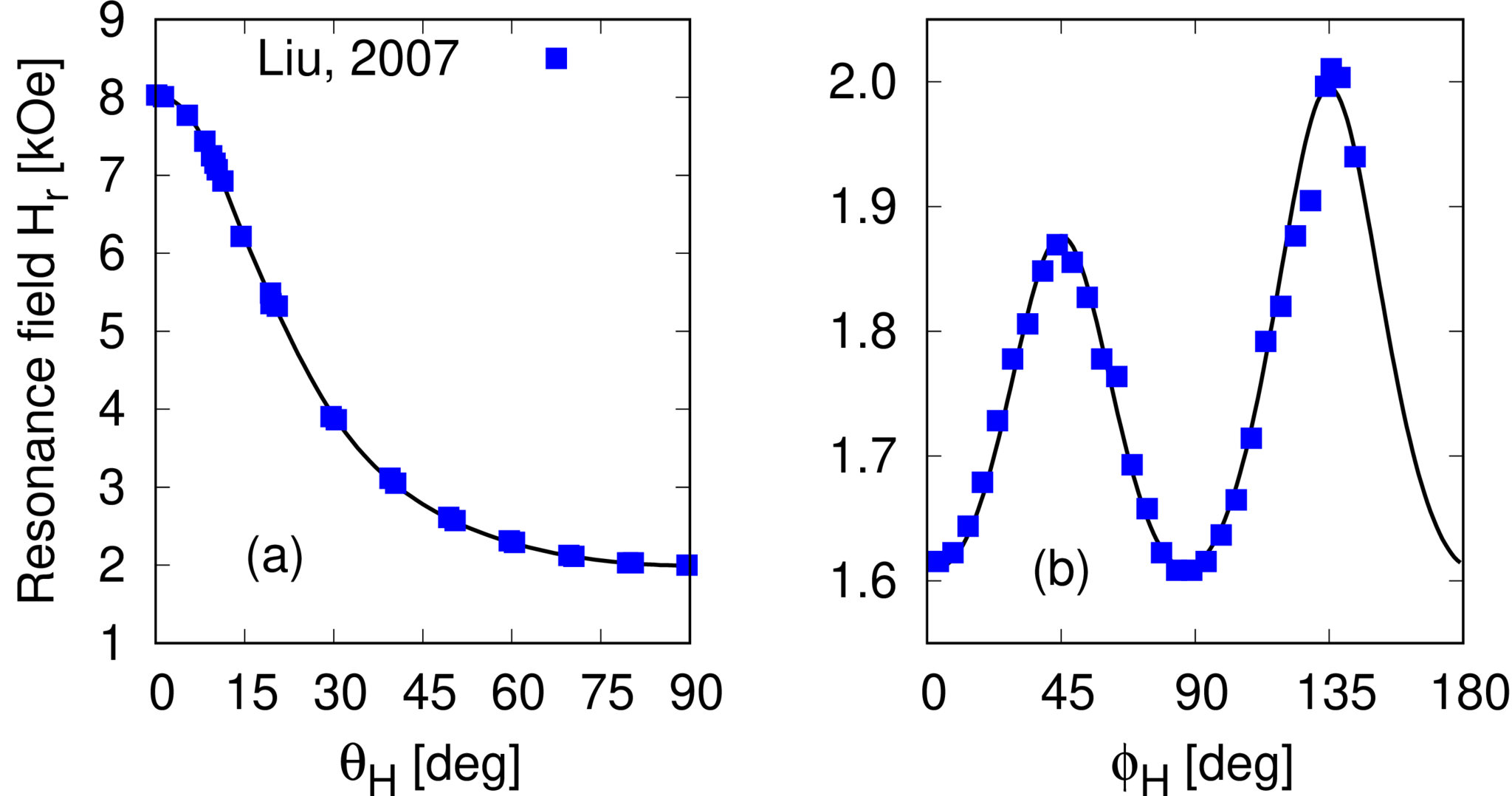}
   \caption{
  Resonance field versus out-of-plane angle $\vartheta_H$ (a)
  and versus in-plane angle $\varphi_H$ (b)
  for (Ga,Mn)As film.
  Squares - Liu measurements [\onlinecite{Liu2007}], 
  black line - solution of Smit-Beljers equation for the vector ${\mathbf h_L}$
  (Eq.~\ref{nv_GaMnAs_fields}).
  Figs~(a) ad (b) correspond to Figs~5 and 6 in Ref.~[\onlinecite{Liu2007}], respectively.
  }
  \label{Liu_resonance}
\end{figure}

\vspace{0.5cm}
\noindent
{\em \bf Numerical example II: Saturation magnetization of the (Ga,Mn)As thin film} ---
In Ref.~[\onlinecite{Tom2018}] we showed, 
analyzing Liu's measurements [\onlinecite{Liu2007}] of resonance fields for the weak 
ferromagnet (Ga,Mn)As,
that in order to properly describe 
the free energy spatial distribution at resonance, one should expand it
up to the fourth order with respect to cubic anisotropy fields.
Here we add to this expansion demagnetizing
fields, and, since the sample was oriented along crystallographical
axes of (Ga,Mn)As we use a diagonal form
od the demagnetizing tensor.
The vector ${\mathbf h_L}$ depends now on eleven  parameters
\begin{equation}
{\mathbf h_L} \equiv  (g, M,\mathcal{N}_x, \mathcal{N}_y, \mathcal{N}_z,  H_{c1},...,H_{c4},H_{[001]},H_{[110]}).
\label{GaMnAs_fields}
\end{equation}
Anisotropy fields are denoted like in Ref.~[\onlinecite{Tom2018}].
The minimization procedure leads to (fields are given in Oe)
\begin{equation}
 \begin{split}
{\mathbf h_L} = (1.984(3), 383.5(2.2), 0.08104(90), 0.06796(95), \\
                    0.8510(20), 76.86(0.42), -539.5(10.0), 42.86(0.92), \\
                    1412(70), 4213(24), 68.08(0.64)).
 \end{split}
\label{nv_GaMnAs_fields}
\end{equation}
The numerical solutions of Eq.~(\ref{SB_eq}) for such an ${\mathbf h_L}$
vector is shown in Fig.~\ref{Liu_resonance}.
They are equivalent to those obtained in Ref.~[\onlinecite{Tom2018}] but now the use 
of the demagnetizing fields made it possible to determine saturation magnetization $M=30.52(0.18)$ [emu/cm$^3$].

\begin{figure}[!htp]
  \includegraphics[width=.8\linewidth]{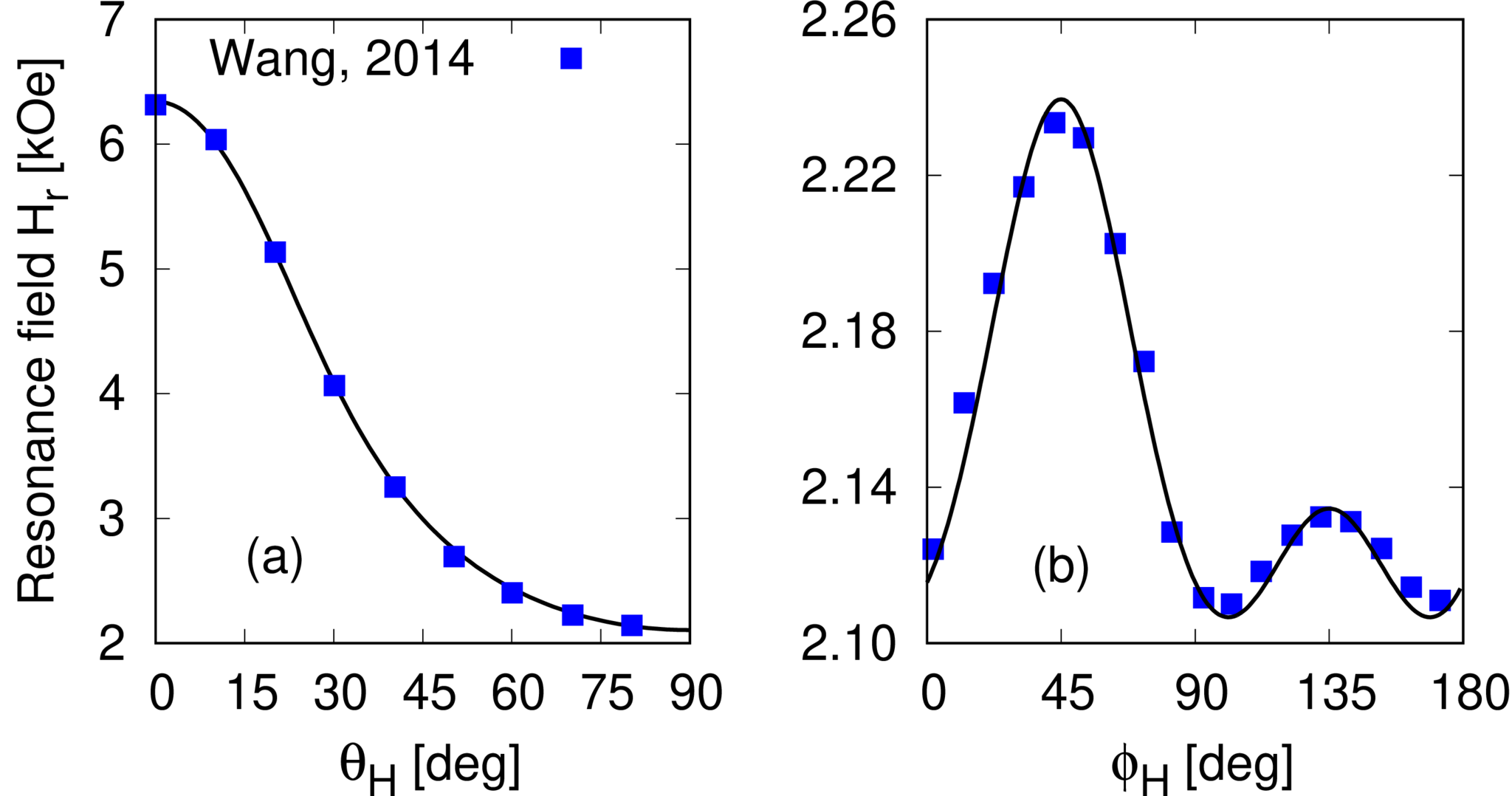}
   \caption{
  Resonance field for 9.8 nm ultrathin YIG film versus out-of-plane angle $\vartheta_H$ (a)
  and in-plane angle $\phi_H$ (b).
  Squares - Wang's measurements [\onlinecite{Wang2014}], 
  black line - solution of Smit-Beljers equation for the vector ${\mathbf h_H}$
  given by Eq.~(\ref{YIG_vector}).
  Figs (a) and (b) correspond to Figs~2(b) and 2(c) in Ref.~[\onlinecite{Wang2014}], respectively .
}
  \label{YIG_fields}
\end{figure}

\noindent
{\em \bf Numerical example III: Saturation magnetization of the YIG ultrathin thin film} ---
The measurements of the resonance field of the 9.8 nm thin film are taken from 
Ref.~[\onlinecite{Wang2014}] and shown in Figs~\ref{YIG_fields}a
and \ref{YIG_fields}b, respectively, as blue squares. 
To determine the anisotropy fields
let us assume that the free energy is given by Eq.~(1) from Ref.~[\onlinecite{Wang2014}]
and that the 
demagnetizing fields, as in the (Ga,Mn)As case,
are calculated using a diagonal demagnetizing tensor. 
Then the vector ${\mathbf h_H}$ depends on nine parameters
\begin{equation}
 \begin{split}
{\mathbf h_H} = (g, M, \mathcal{N}_x, \mathcal{N}_y, \mathcal{N}_z, H_{2\perp}, H_{4\perp}, H_{2\parallel}, H_{4\parallel}).
\end{split}
\label{YIG_vector}
\end{equation}
Factor $g$ = 2 for YIG [\onlinecite{Stancil}], while saturation magnetization $M$ = 1851 [Oe] 
was determined for this sample using a~vibrating sample magnetometer [\onlinecite{Wang2014}].

The components of the vector ${\mathbf h_H}$ 
corresponding to the constant curvature of free energy, i.e.,
for the sample at resonance,
are collected in Table~\ref{Hammel_Fields} for four cases. In the first case, 
we take $g$ and $M$ values as measured experimentally
and find remaining ones.
In the second case we fix only experimental value of $M$, in the third case - only value of $g$, 
and finally, we treat both $g$ and $M$ as unknown (last row of in Table~\ref{Hammel_Fields}). 
In the last column of Table~\ref{Hammel_Fields}
the error function $E_{RMS}^{1}({\mathbf h_H} )$ is given, as defined in Ref.~[\onlinecite{Tom2018}].
Its value informs us about the quality of the prediction of new experimental results 
of resonance fields calculated from a given free energy formula with 
specific values (row of Table~\ref{Hammel_Fields}) of anisotropy fields.
We see that analysis of FMR measurements of the YIG ultrathin film gives worse 
results when we treat measured $g$ and $M$ values as known.
This points out that the form of free energy used for this analysis is not chosen optimally, 
because it was not possible to reproduce the experimental 
values $g$ and $M$ with satisfactory accuracy.
Thus the presented method may also serve as a test for the correctness 
of the assumed free energy form.
Note also, that although the investigated film was ultrathin, 
we obtained non-zero values of the demagnetization fields in the 
(001) plane.
The dependencies of resonance fields on angles $\vartheta_H$ and $\varphi_H$ 
are shown, for ${\mathbf h_H}$ from the first row of Table~\ref{Hammel_Fields}, in Fig.~\ref{YIG_fields}.
\begin{table*}
\caption{
Comparison of $g$-factor, saturation magnetization $M$ and 
and anisotropy field values
found in Ref.~[\onlinecite{Wang2014}] with those
obtained by the method described
in the text.
The errors (in brackets) refer to the last significant digits.
}
\begin{tabular}{ccccccccccc}
\hline 
  &   $g$     &   $M$[Oe]   &$\mathcal{N}_x$& $ \mathcal{N}_y$&  $\mathcal{N}_z$ & $H_{2\perp}$[Oe]& $H_{4\perp}$[Oe]
  & $H_{2\parallel}$[Oe]&$H_{4\parallel}$[Oe]&$E_{RMS}^{1}({\mathbf h_H} )$\\
\hline
  &  2.000\footnote{Measurement result, see Apendix A of Ref.~[\onlinecite{Stancil}].} 
  & 1851\footnote{Measurement result, Ref.~[\onlinecite{Wang2014}].} & 0.08429(10) & 0.08721(11) &  0.8285(10)  & -497.7(2.0)  &   372.3(1.1)   &  62.23(0.27)   &   49.02(0.22)  &   0.815\\
  &  2.014(1) & 1851$^\textrm{b}$ & 0.08795(17) & 0.08985(10) &  0.8222(15)  &  -466.5(1.5)  &   276.9(0.9)   &  61.83(0.23)   &   47.27(0.18)  &   0.691\\
  &  2.000$^\textrm{a}$ & 1800(11) & 0.08102(15) & 0.08408(11) &  0.8349(16)  &  -538.9(1.7)  &   372.6(1.0)   &  62.16(0.11)   &   49.05(0.16)  &   0.815\\
  &  2.013(1) & 1772(12) & 0.08263(8) & 0.08467(13) &  0.8327(13)  &  -526.2(1.6)  &   277.5(0.9)   &  61.82(0.09)   &   47.42(0.18)  &   0.691\\
 Ref. [\onlinecite{Wang2014}]  &   & 1851(37) &  &              &  & -1253(25) & & 60.4(1.2) & 42.0(1.2) \\
\hline
\end{tabular}
\label{Hammel_Fields}
\end{table*}

\vspace{0.5cm}
\noindent
{\em \bf Conclusion} --- 
We presented in this article
one more approach to the interpretation of the classical FMR experimental data. 
It is based on the analysis of the curvature of the spatial dependence of free energy
of the sample at resonance and makes it 
possible, 
after assuming the functional form of free energy,
to determine (with an accuracy of 0.5-4\% in the presented numerical examples) the sample saturation magnetization
and, consequently, the spatial dependence of 
magnetocrystalline energy stored in the sample in a single FMR experiment.
The method presented here can also be a test
for the correctness of the assumed form of free energy
of the sample at resonance.
This may be important for people working in the field of magnetic resonance.

\vspace{0.5cm}
\noindent
{\bf Acknowledgment} --- This study is a~part of the~project financed by Narodowe
Centrum Nauki (National Science Centre of Poland)
No.~DEC-2013/08/M/ST3/00967. Numerical calculations were performed at Pozna\'n Supercomputing and
Networking Center under Grant No.~284.
\bibliographystyle{apsrev4-1}
\bibliography{Ms_Curvature.bib}

\end{document}